\documentclass[11pt]{article}
\usepackage{moriond,epsfig} 

\def\be{\begin{equation}}
\def\ee{\end{equation}}
\def\bea{\begin{eqnarray}}
\def\eea{\end{eqnarray}}

\begin{document}
\vspace*{4cm}
\title{HIGGS SEARCHES AT THE LHC}

\author{ T. VICKEY\\ On behalf of the ATLAS and CMS Collaborations}

\address{Department of Physics, University of Wisconsin-Madison,\\ 1150 University Avenue
Madison, WI 53706-1390, USA}

\maketitle\abstracts{These proceedings summarize the sensitivity for the ATLAS and CMS experiments at the LHC to discover a Standard Model Higgs boson with relatively low integrated luminosity per experiment.  A brief discussion on the expected performance from these experiments in searches for one or more of the Higgs bosons from the minimal version of the supersymmetric theories is also included.}

\section{Introduction}
The primary objective of the LHC is to elucidate the mechanism responsible for electroweak symmetry breaking.  In the context of the Standard Model (SM) the assumption of one doublet of scalar fields gives rise to a scalar particle known as the Higgs boson.\cite{peter_higgs,cs_lhc}  The Higgs mass is not predicated by theory and, to date, direct experimental searches for the Higgs have put a lower limit on its mass at $M_H > 114.4\mbox{ GeV}/c^2\mbox{ @ } 95\%$ CL.\cite{lep}  A preferred value for the Higgs mass, derived by fitting precision electroweak data,\cite{lepewwg} is currently $M_H = 87^{+36}_{-27}\mbox{ GeV}/c^2$ with an upper bound of $160\mbox{ GeV}/c^2\mbox{ @ } 95\%$ CL. 

Both the ATLAS and CMS experiments at the LHC, scheduled for proton-proton collision data-taking  beginning Summer 2008, have been designed to search for the Higgs over a wide mass range.\cite{atlas_tdr,cms_tdr}  These proceedings summarize the sensitivity for each experiment to discover a SM Higgs boson with relatively low integrated luminosity per experiment ($1 - 30$ fb$^{-1}$) as well as recent developments that have enhanced this sensitivity.  A brief discussion on the sensitivity for these experiments to discover one or more of the Higgs bosons from the minimal version of the supersymmetric theories~\cite{susy} (MSSM) is also included.

\section{Standard Model Higgs Production at the LHC}
\begin{figure}[ht]
\begin{minipage}[b]{0.5\linewidth}
\centering
\includegraphics[scale=1.5]{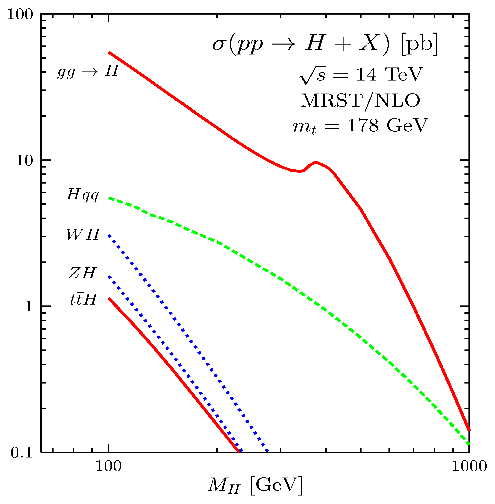}
\caption{Dominant Standard Model Higgs production cross-sections at the LHC.$^2$}
\label{fig:higgs_production}
\end{minipage}
\hspace{0.5cm}
\begin{minipage}[b]{0.5\linewidth}
\centering
\includegraphics[scale=1.53]{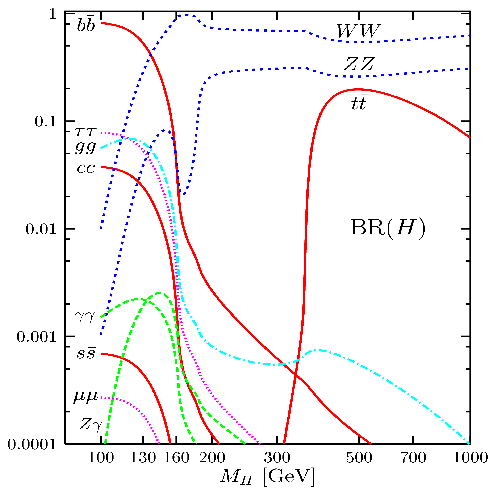}
\caption{Branching ratios for Standard Model Higgs decays.$^2$}
\label{fig:higgs_decay}
\end{minipage}
\end{figure}

The SM Higgs production cross-sections at the LHC (to NLO), as a function of Higgs mass, are shown in Figure~\ref{fig:higgs_production}.  The dominant production mechanism for SM Higgs boson production, which proceeds via a top-quark loop, is the {\em Gluon-Gluon Fusion} mode ($gg\rightarrow H$).  The {\em Vector Boson Fusion} (VBF) process ($qq\rightarrow Hqq$) is the second-most dominant production mode at the LHC.  {\em Associated Production} modes, where the Higgs is produced via $q\overline{q}'\rightarrow HW$, $q\overline{q}\rightarrow HZ$ and $gg,q\overline{q}\rightarrow t\overline{t}H$, have much smaller cross-sections.  The presence of a $W$, $Z$ or top-quark alongside the Higgs, or high-$p_T$ high-$\eta$ jets from VBF, allow for triggering on events with Higgs in invisible final states.

\section{Higgs Discovery Final States}
The final states most suitable for discovery at the LHC vary depending on the branching ratios (shown in Figure~\ref{fig:higgs_decay}), which are a function of the Higgs mass, and the relevant backgrounds.  For $M_H < 2M_W$ the dominant decay mode is through $b\overline{b}$; however, due to the enormous QCD background, this channel is only considered in the $t\overline{t}H$ final state where handles exist for the rejection of this background.  The $\gamma\gamma$ final state, which appears when the Higgs decays via bottom, top and $W$ loops, has a small branching fraction but excellent $\gamma/\mbox{jet}$ separation and $\gamma$ resolution help to make this a very significant channel.  $H\rightarrow\tau^+\tau^-$ is accessible through the VBF modes, where the two struck quarks appear as high-$p_T$ jets in the very forward (high-$\eta$) and opposite regions of the detectors.

If the Higgs mass is large enough to make the $WW$ and $ZZ$ modes kinematically accessible, the $H\rightarrow WW^{(*)}$ final-states are powerful over a very large mass range ($WW$ accounts for $\sim$95\% of the branching ratio at $M_H \sim$160$\mbox{ GeV}/c^2$), as is the $H\rightarrow ZZ^{(*)}\rightarrow 4l$ final state--the latter of which is commonly referred to as the ``Golden Mode'' as with four leptons in the final state the signal is easy to trigger on and allows for full reconstruction of the Higgs mass.

Both ATLAS and CMS have conducted extensive fully-simulated {\tt GEANT}-based~\cite{geant} Monte Carlo studies to determine the experimental viability of all of these channels.  A few of these signatures are highlighted below; a more comprehensive and complete account can be found elsewhere.\cite{atlas_tdr,cms_tdr}

\subsection{$H\rightarrow\gamma\gamma$}
Despite the small branching ratio, $H\rightarrow\gamma\gamma$ remains a very attractive channel for $M_H < 140\mbox{ GeV}/c^2$.  Genuine photon pairs from $q\overline{q}\rightarrow\gamma\gamma$, $gg\rightarrow\gamma\gamma$ and quark bremsstrahlung comprise the irreducible background, while jet-jet and $\gamma$-jet events, where one or more jets are misidentified as photons, make up the reducible background.  $Z\rightarrow ee$ events, with both electrons misidentified as photons, can be reduced using electron/photon separation techniques.  The sensitivity of this channel is similar for both experiments; the high-granularity liquid Argon calorimeter of ATLAS is capable of determining the primary vertex with great precision, while CMS has a superior energy resolution.  Studies conducted by both experiments consider the signal and background to NLO.  Both experiments have looked beyond a simple cut-based analysis and enhanced the signal significance by $\sim$35\%.  For $M_H = 130\mbox{ GeV}/c^2$, and an integrated luminosity of 30 fb$^{-1}$, the significance~\cite{cms_tdr} is just above $8\sigma$.

\subsection{$H\rightarrow ZZ^{(*)}\rightarrow 4l$ ($4e$, $4\mu$, and $2e2\mu$)}
At $M_H > 130\mbox{ GeV}/c^2$, the 4-lepton channels gain in importance on account of the precise energy reconstruction of both ATLAS and CMS for electrons and muons.  The dominant backgrounds for these channels are $ZZ^{(*)}$, $Zb\overline{b}$ and $t\overline{t}$ production.  Through the use of impact parameter and lepton isolation requirements the latter two can be significantly reduced.  The $q\overline{q}$ component of the $ZZ^{(*)}$ background is known at NLO, however due to the lack of a Monte Carlo generator~\footnote{The current efforts of those working on the {\tt gg2ZZ} generator were recently brought to the attention of the author.} for $gg\rightarrow ZZ^{(*)}$, typically the contribution from this process is added as 30\% of the LO $q\overline{q}\rightarrow ZZ^{(*)}$.  Collectively, the significance for these channels is more than $5\sigma$ for 30 fb$^{-1}$ of data.

\subsection{$H\rightarrow WW^{(*)}\rightarrow l\nu l\nu$ ($l=e$, $\mu$)}
As the branching ratio for a SM Higgs decaying to $WW$ is more than $95\%$ at $\sim$$160\mbox{ GeV}/c^2$, this is the most significant channel at that mass point.  Unlike other channels, in the $H\rightarrow WW\rightarrow l\nu l\nu$ final state full mass reconstruction is not possible and the analysis is essentially reduced to a counting experiment; therefore an accurate background estimate is critical.  The dominant backgrounds for this analysis are $WW$ and $t\overline{t}$ production.  The former can be suppressed by exploiting spin correlations between the two leptons while the latter has been shown to be suppressed significantly by a jet veto.  Using NLO cross-sections and conservative estimates for the effect of systematic uncertainties, a significance of around $5\sigma$ for $M_H = 165\mbox{ GeV}/c^2$ using an integrated luminosity of $\sim$1 fb$^{-1}$ is estimated.

\subsection{$H\rightarrow \tau^+\tau^-$}
The distinct experimental signature of Higgs production via VBF, with jets from the ``struck quarks'' at high-$\eta$ and Higgs decay products (but very little else) in the central region is a great asset for channels like $H\rightarrow \tau^+\tau^-$.  ATLAS and CMS now both consider three final states, thus covering all combinations of leptonically- and hadronically-decaying taus.  Triggering on the fully hadronic mode by using combinations of low-$p_T$ tau and other triggers (e.g., missing transverse energy or forward jets) are currently under investigation.  Despite the presence of multiple neutrinos in the final-state, mass reconstruction can typically be done via the collinear approximation where the tau decay daughters are assumed to be in the same direction as their parent.  The resolution on the reconstructed mass ($\sim$10$\mbox{ GeV}/c^{2}$ for $M_H = 150\mbox{ GeV}/c^{2}$) is mainly affected by the missing energy resolution.  Data-driven methods for understanding the dominant backgrounds ($Z+\mbox{jets}$, QCD and $t\overline{t}$) have been investigated.

\section{Summary of Higgs Discovery Potential}
\begin{figure}[ht]
\begin{minipage}[b]{0.5\linewidth}
\centering
\includegraphics[scale=0.42]{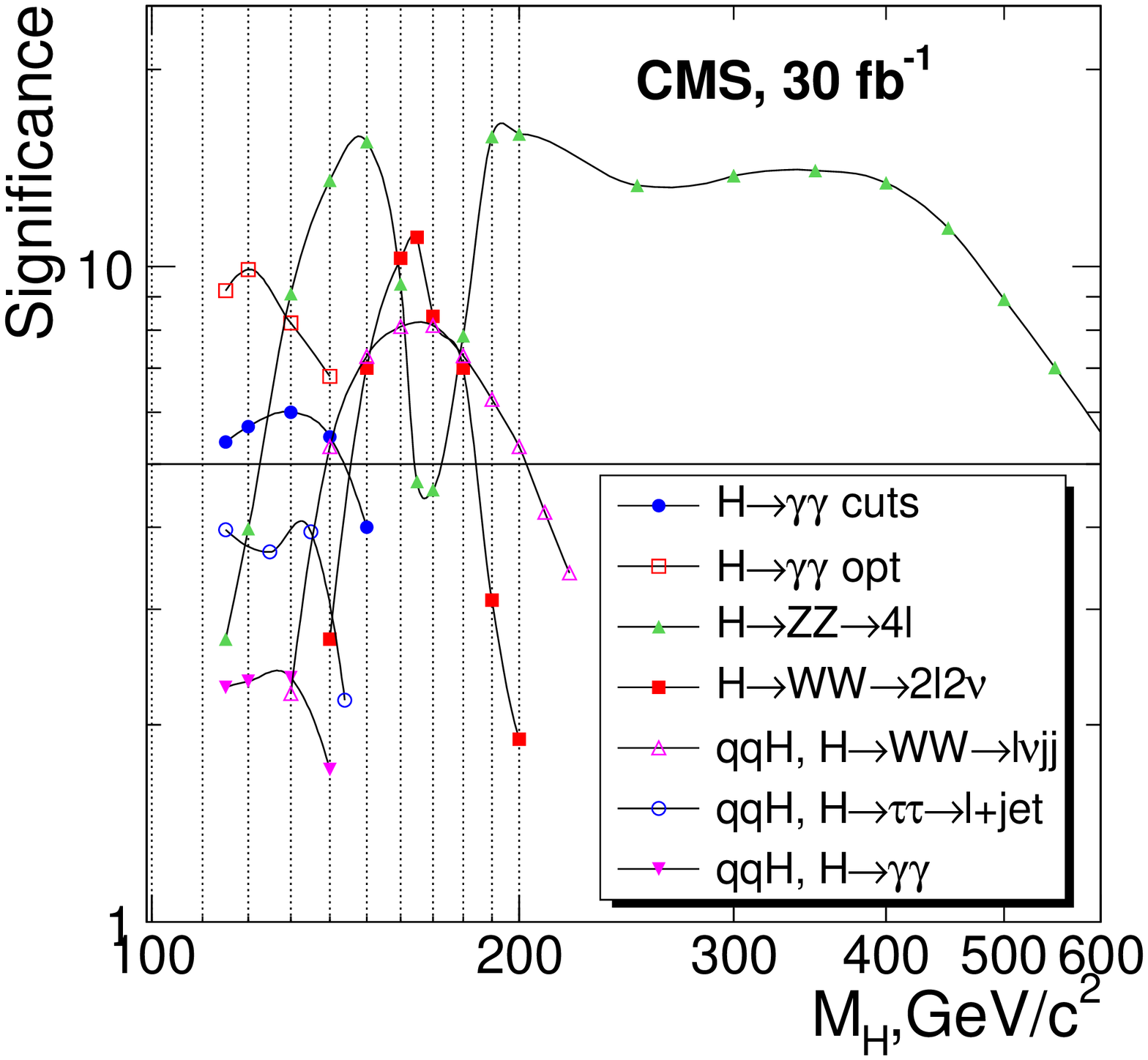}
\caption{The discovery potential at CMS$^6$ for Standard Model Higgs boson searches, as obtained using NLO cross-sections, for 30 fb$^{-1}$.}
\label{fig:cms_sensitivity}
\end{minipage}
\hspace{0.5cm}
\begin{minipage}[b]{0.5\linewidth}
\centering
\includegraphics[scale=0.42]{ATLAS_TDR_chapter19_fig3.epsi}
\caption{ATLAS sensitivity$^5$ for the discovery of MSSM Higgs bosons (minimal mixing scenario).  The $5\sigma$ discovery contour curves are shown in the $m_A - \tan\beta$ plane for 30 fb$^{-1}$.  Performance with CMS is similar.$^6$}
\label{fig:atlas_mssm_sensitivity}
\end{minipage}
\end{figure}

The expected significance in 30 fb$^{-1}$, for various final states as a function of SM Higgs mass, is summarized in Figure~\ref{fig:cms_sensitivity}.  The discovery potential at ATLAS and CMS is quite similar.

Discovery prospects for the detection of MSSM Higgses ($A$, $h$, $H$ and $H^{\pm}$) have also been evaluated.\cite{atlas_tdr,cms_tdr}  At tree-level, all Higgs masses and couplings can be expressed in terms of $m_A$ and $\tan\beta$.  The complete region of the $m_A - \tan\beta$ parameter space ($m_A = 50 - 500\mbox{ GeV}/c^2$ and $\tan\beta = 1 - 50$) should be accessible to the LHC experiments.  The sensitivity for the discovery of MSSM Higgses, in the minimal mixing scenario for 30 fb$^{-1}$ of data, is summarized in Figure~\ref{fig:atlas_mssm_sensitivity}.

\section*{Acknowledgments}
The author would like to thank the ATLAS and CMS Collaborations, in particular L. Fayard, C. Mariotti, A. Nisati and Y. Sirois, for their advice and support.  The author was supported in part by the United States Department of Energy through grant number DE-FG02-95ER40896.

\end{document}